\documentclass[trackchanges, twocolumn]{aastex701}

\usepackage{xcolor}
\usepackage{amsmath}
\usepackage{hyperref}
\usepackage{multirow}
\usepackage{xspace}

\begin{document}

% \title{Multimessenger Study of Supermassive Black Hole Accretion Rates for $0.2 \leq z \leq 6.2$}
% \title{Multimessenger Study of Supermassive Black Hole Eddington Fractions for $0.2 < z < 6.2$}
% \title{Finally, GWB and QLF united at last!}
\title{A Unified Model for the Gravitational Wave Background and Quasar Luminosity Functions}
% \title{A Model to Unify the Gravitational Wave Background With Quasar Luminosity Functions}
% \title{The Gravitational Wave Background is Consistent With Quasar Luminosity Functions}
% \title{The Gravitational Wave Background Agrees With Quasar Luminosity Functions}
% \title{One Model to Rule Them All: The Gravitational Wave Background Agrees With Quasar Luminosity Functions}
% \title{Astronomers hate this one simple trick to explain both the gravitational wave background and quasar luminosity functions}

\author[0000-0002-9710-6527]{Cayenne Matt}
\affiliation{Department of Astronomy and Astrophysics, University of Michigan, Ann Arbor, MI 48109, USA}
\email[show]{cayenne@umich.edu}

\author[0000-0002-1146-0198]{Kayhan G\"{u}ltekin}
\affiliation{Department of Astronomy and Astrophysics, University of Michigan, Ann Arbor, MI 48109, USA}
\email{kayhan@umich.edu}

\author[0000-0003-1468-9526]{Ryan C. Hickox}
\affiliation{Department of Physics and Astronomy, Dartmouth College, 6127 Wilder Laboratory, Hanover, NH 03755, USA}
\email{Ryan.C.Hickox@dartmouth.edu}

\newcommand{\msigma}{$M_\mathrm{BH}$--$\sigma$}

\newcommand{\mmb}{$M_\mathrm{BH}$--$M_\mathrm{bulge}$}

\newcommand{\lamedd}{$\lambda_\mathrm{Edd}$}

\newcommand{\comm}{\textcolor{cyan}}

\newcommand{\mone}{$\mathrm{M}_\mathrm{Fiducial}$}
\newcommand{\mtwo}{$\mathrm{M}_\mathrm{GWB}$}

\newcommand{\gone}{$\gamma_1$}
\newcommand{\gtwo}{$\gamma_2$}
\newcommand{\lbk}{$\lambda_\mathrm{break}$}
\newcommand{\factive}{$\mathrm{F}_\mathrm{norm}$}

\newcommand{\twoline}[1]{\multirow{2}{*}{#1}}

\begin{abstract}
The supermassive black hole mass function (BHMF) is a necessary ingredient in our understanding of how BHs grow and evolve with their host galaxies. Estimations of the BHMF based on observations of local BHs, quasars, and gravitational wave data yield inconsistent results, leading to large uncertainties in, e.g., BH accretion and seeding models. The nanohertz gravitational wave background (GWB) has been a powerful tool to provide constraints on the BHMF across cosmic time. We take posterior BHMFs that describe the GWB and fit them to bolometric quasar luminosity functions (QLF) across $0.2 \leq z \leq 6.2$ to constrain the Eddington ratio distribution function (ERDF). Compared to a fiducial BHMF, based on galaxy and a local black hole mass scaling relation, the BHMFs that match the GWB can reproduce the observed QLFs with lower Eddington fractions and without requiring significant super-Eddington accretion. We use a double power-law form of the ERDF and find that the fits using the GWB-inferred BHMF have systematically lower break-points and marginally steeper slopes in the high-Eddington fraction portion of the function. Furthermore, the best-fit QLFs using the GWB-inferred BHMFs provide better fits to the high-luminosity end of the QLF data than the fiducial model. For both models the ERDF shows significant redshift evolution of the break-point which reaches a maximum around $z \sim 2$. Overall, our prescription can produce realistic QLFs from BHMFs and our GWB-based model is consistent with electromagnetic and gravitational wave data sets.
\end{abstract}

% \keywords{\uat{Galaxies}{573} --- \uat{Cosmology}{343} --- \uat{High Energy astrophysics}{739} --- \uat{Interstellar medium}{847} --- \uat{Stellar astronomy}{1583} --- \uat{Solar physics}{1476}}


\section{Introduction}

Inside nearly every massive galaxy grows a supermassive black hole \citep[SMBH, ][]{Kormendy_Ho_2013}. Throughout their lifetime, these SMBHs will undergo phases of accretion activity and quiescence. When the energy released from accretion is large enough, these SMBHs can be observed as active galactic nuclei (AGN, also ``quasars''). These AGN can have luminosities of $10^{42} \lesssim  L_\mathrm{AGN} \lesssim 10^{48} \;\mathrm{erg} \; \mathrm{s}^{-1}$ and can outshine their host galaxies, making them observable to incredibly high redshifts thus enabling studies of otherwise elusive SMBHs in the very early stages of growth \citep{Padovani_2017, Alexander_2012}. The majority of AGN are observed in the X-ray due to low attenuation at these wavelengths and because X-ray sources are relatively rare and are thus an excellent indicator of accretion. The bolometric quasar luminosity function (QLF), on the other hand, can give insight into the AGN population across a wider range of masses and accretion rates \citep{Shen_2020}.

One of the most fundamental properties of SMBH populations is the mass distribution, i.e., the SMBH mass function (BHMF). Knowing the relative numbers of SMBH masses can teach us about the population of early SMBH seeds as well as their growth history and coevolution with their host galaxies \citep{Greene_2020, Kormendy_Ho_2013}. While direct SMBH mass measurements are infeasible outside the local universe, secondary methods of inferring SMBH masses can be quite powerful.

Previous works have leveraged AGN surveys to place constraints on the the BHMF. From these studies we have gained insight into the formation and growth of SMBHs \citep[e.g.,][]{Hopkins_2007e, Shen_2020, Zou_2024_bhmf} as well as enabling the study of galactic environments on both large and small scales \citep[e.g.,][]{Menard_2010, Rudie_2012, Johnson_2017, Mishra_2024}. From these studies it has become clear that SMBHs play a large role in the evolution of their host galaxies and that the mass of both objects are closely intertwined and sensitive to the energetic feedback of their growth \citep{Kormendy_Ho_2013}. There exist, however, several tensions between the BHMFs inferred from AGN and other data-driven methods. Constraints on the BHMF from AGN observations often rely on scaling relations and assumptions about the accretion rate, Compton thick fraction, radiative efficiency, and the fraction of all SMBHs that are active and observable and survey completeness fractions. Even when normalizing measured accretion rates to the well-measured host stellar mass, there is still an underlying assumption that there is one global, unchanging, relation between host stellar mass and SMBH mass \citep[i.e. \mmb][]{Kormendy_Ho_2013, McConnell_2013, Aird_2018, Zou_2024_bhar}.

% Some AGN studies have focused on the \mmb\ relation itself. For example, \citet{Caplar_2015, Caplar_2018} found that the evolution in the characteristic luminosity of the QLF, $L_*$, may be due to either an evolution of the ERDF or in the SMBH--galaxy mass ratio, $M_\mathrm{BH} / M_\star$ or a combination of the two. They concluded that a strong redshift evolution $M_\mathrm{BH} / M_\star \propto (1 + z)^{1.5-2}$ can explain both the $L_*$ evolution as well as the apparent downsizing. Alternatively, others have found evidence to support a non-evolving and/or relatively low $M_\mathrm{BH} / M_\star$ ratio \citep{Shankar_2016, Zou_2024_bhmf}.

Under the assumption that all galaxies host a SMBH with a mass that is some constant fraction of the host bulge mass \citep[e.g., \mmb][]{Kormendy_Ho_2013, McConnell_2013}, one can also predict the SMBH population from observations of galaxies and local scaling relations. Compared to AGN-based results, this method of predicting SMBH mass functions can produce mass densities (local and at higher redshift) that are greater by a factor of 2 or more \citep[e.g., ][]{Liepold_2024, Shankar_2016, Aird_2013}. This tension can be alleviated if the assumed \mmb\ relation has a relatively low normalization and/or if a significant fraction of, especially the most massive, SMBHs are accreting at super-Eddington rates \citep{Shankar_2016, Aird_2013}.

Recently, four pulsar timing array (PTA) collaborations announced strong evidence of a common quadrupolar signal indicative of a gravitational wave background \citep[GWB, ][]{Agazie_2023, Antoniadis_2023, Reardon_2023, Xu_2023}. Under the assumption that the GWB is produced by SMBH binaries, it is possible to predict the strain amplitude based on the electromagnetic-observation-based SMBH mass function. Surprisingly, the observed GWB amplitude is a factor of 2--3 times higher than expected from electromagnetic constraints. This increases the tension with AGN-based BHMFs and brings our assumptions for SMBH--galaxy growth and evolution into question \citep{AgazieBHB_2023}.

It has now been just over three years since the initial announcement. In that time, numerous studies have independently concluded that the high GWB amplitude requires an elevated number density of the most massive SMBHs by up to a factor of 10 relative to predictions from local scaling relations \citep{AgazieBHB_2023, Chen_2024, Sato-Polito_2024, Liepold_2024, Matt_2026a, Matt_2026b}. There are several solutions that have been offered to help explain this numerous population of massive SMBHs. \citet{Liepold_2024} found that the most massive galaxies $\log_{10} M_\star/M_\odot \geq 11.5$ (hosts of the most massive SMBHs) may be severely under counted within the local volume in cosmological surveys. \citet{Chen_2024} and \citet{Matt_2026a, Matt_2026b} found that the black hole--galaxy mass ratio may be evolving positively with redshift. \citet{Siwek_2024} and \citet{Comerford_2025} proposed that SMBH binaries may experience significant accretion post galaxy merger which can significantly boost both the chirp mass of the binary, but also the mass ratio, both producing an increased GWB amplitude. Regardless of which particular model is used, the result is broadly the same; the GWB indicates a significant population of extremely massive SMBHs ($\log_{10} M_\mathrm{BH}/M_\odot > 9$). If the SMBH population is indeed significantly inflated, the implications for AGN studies are profound.

The high number density of SMBHs implied by the GWB indicates that, for the same AGN luminosity density, the average accretion rates across the SMBH population must be lower than previously assumed or that the AGN fraction is over estimated. There are several studies that have investigated AGN accretion in the context of the GWB. For example, \citet{Tillman_2026} used cosmological simulations to model SMBH growth and compared how different accretion models produce SMBH populations with notably different GWB amplitudes. They concluded that, to grow SMBHs to the masses necessary to produce the GWB amplitude, SMBH growth must occur earlier than current models predict or accretion must be more efficient. Also using cosmological simulations \citet{Bonoli_2025} compared the GWB amplitude from their SMBH populations to PTA data and also considered how well the AGN luminosity function produced by their accretion models reproduced QLF data. The models that best described all datasets required either significant rates of heavy BH seed formation or large numbers of light seeds that experience repeated phases of super-Eddington accretion.

Starting from the observed QLF \citet{Kis-Toth_2025} were able to reproduce the amplitude reported by \citet{Agazie_2023}. To do so they assumed that every AGN in the QLF is powered by an SMBH binary, and found that a typical Eddington ratio of $\sim 25\%$ and a typical lifetime of $\sim 3\times10^{7}\;\mathrm{yr}$ can reconcile the two datasets. The fraction of AGN hosting SMBH binaries is highly uncertain. By combining the GWB from \citet{Agazie_2023} with periodic quasar candidates from Catalina Real-time Transient Survey, \citet{Casey-Clyde_2025} found that the fraction of AGN currently hosting a PTA-band binary is at most $\lesssim 5\%$, implying that quasars are roughly five times as likely as random massive galaxies to host a binary. While there has never been a confirmed gravitationally-bound SMBH binary, numerous dual AGN have been found, some at remarkably small separations \citep[e.g., $\sim 230$\,pc;][]{Koss_2023}.

In this paper we use SMBH mass functions inferred from the GWB to construct QLFs. We fit these QLFs to the observed bolometric QLF across $0.2 \leq z \leq 6.2$ to place constraints on the Eddington ratio distribution function. We describe our data and methods in Section \ref{sec:data_and_model}. In Section \ref{sec:results} we present the results of our models and we compare to literature measurements of the ERDF. We discuss the broader implications of our findings in Section \ref{sec:discussion} and summarize this paper in Section \ref{sec:summary}. We use WMAP9 cosmology throughout this work \citep{wmap9}.

\section{Data and Model}\label{sec:data_and_model}

\begin{figure*}[ht!]\centering
    \includegraphics[width=\textwidth, keepaspectratio]{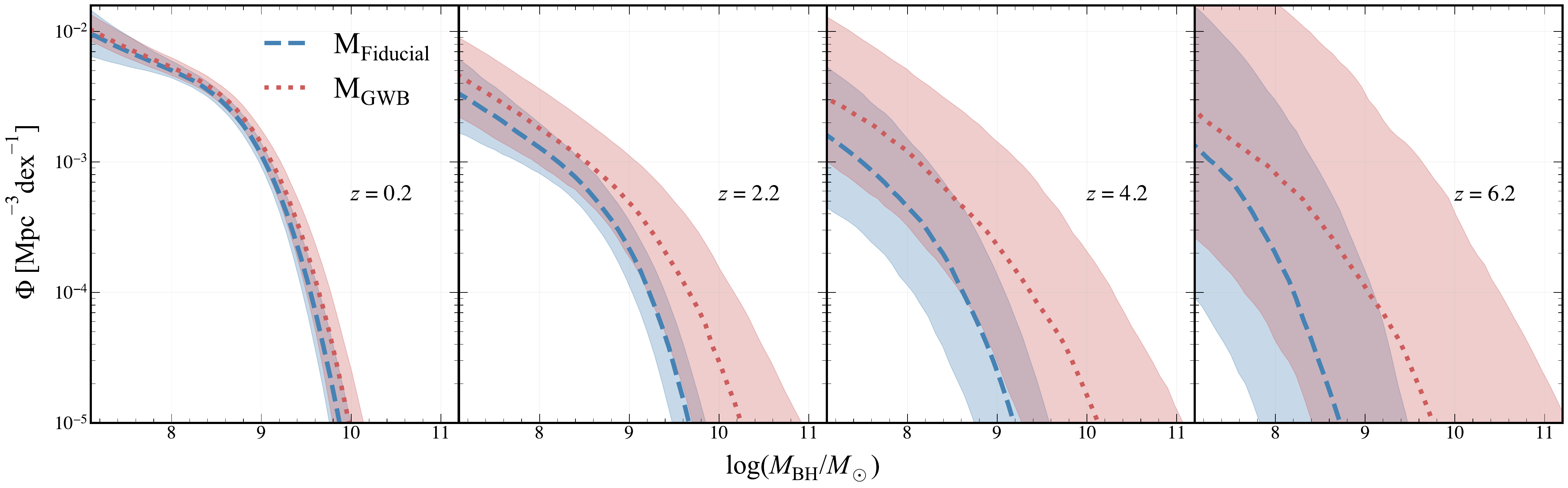}
    \caption{The BHMFs for the two models we consider in this work. The blue dashed line is the fiducial model (\mone) made from a convolution between the GSMF from \citet{Leja_2020} and the local \mmb\ relation from \citet{Kormendy_Ho_2013}. The red dotted line represents our \mtwo\ model which is inferred from the GWB \citep[referred to as Le11ev in][also see their Figure 7]{Matt_2026a}. The shaded regions represent the 16th and 84th confidence intervals. We show the highest, lowest, and two intermediate redshifts considered in this work. The two BHMFs are identical for $z = 0.0$ by design and \mtwo\ includes positive normalization evolution in the \mmb\ relation which increases the BHMF number density to be consistent with the GWB and drives the increasing difference between the two BHMFs with redshift.}
    \label{fig:bhmfs}
\end{figure*}

For this work, we map SMBH mass functions (BHMFs) onto QLFs by convolving them with an Eddington ratio distribution function (ERDF, $P(\lambda_\mathrm{Edd})$). We assume a broken power-law form to the ERDF defined as a function of linear \lamedd\

\begin{equation}
    P(\lambda_\mathrm{Edd}) = \frac{1}{\mathcal{N}}
    \begin{cases}
        \left( \frac{\lambda_\mathrm{Edd}}{\lambda_\mathrm{break}} \right)^{\gamma_1}, & \lambda_\mathrm{Edd} < \lambda_\mathrm{break} \\
        \left( \frac{\lambda_\mathrm{Edd}}{\lambda_\mathrm{break}} \right)^{\gamma_2}, & \lambda_\mathrm{Edd} \ge \lambda_\mathrm{break}
    \end{cases}
    \label{eqn:plambda}
\end{equation}
where \gone\ and \gtwo\ are the slopes of the power law at low and high values of \lamedd. The point where the function transitions between the two slopes is \lbk. $\mathcal{N}$ is the normalization such that $P(\lambda)$ integrates to 1 across our chosen range of $-4 < \log_{10}(\lambda) < 1$ (we set the upper limit to 0 for our Eddington-limited case discussed in Section \ref{sec:eddlim}).

Our final framework\footnote{\url{https://github.com/CayenneMatt/gwb_analysis/tree/main}} has four free parameters, the upper and lower slopes of the ERDF, the break-point or knee of the ERDF, and a number density normalization factor, \factive. \factive\ is a multiplicative factor we apply in the convolution step and functionally serves as the theoretical fraction of all SMBHs that are accreting with $\log_{10}(\lambda_\mathrm{Edd}) \geq -4$. This parameter is related, but not equivalent to, the active fraction, we discuss this further in Section \ref{sec:results}. The full convolution is given by

\begin{multline}
    \Phi_L(\log_{10} L_\mathrm{bol}, z) =
    \int \Phi_\mathrm{BH}\big(\log_{10} M_\mathrm{Edd},z\big)\; \times \\
    P\big(\lambda_\mathrm{Edd}\big)\;
    F_\mathrm{norm}\;d\log_{10}\lambda_\mathrm{Edd}.
\end{multline}
where $\Phi_L(\log_{10} L_\mathrm{bol}, z)$ is the QLF per unit log-luminosity per volume at a given redshift $z$, $\Phi_\mathrm{BH}\big(\log_{10} M_\mathrm{Edd},z\big)$ is the BHMF per unit log-Eddington mass per volume at a given redshift, $P\big(\lambda_\mathrm{Edd}\big)$ is ERDF from Equation \ref{eqn:plambda} and \factive\ is the number density normalization factor.

We fit our model QLFs to the bolometric QLF data from \citet{Shen_2020} for $0.2 \leq z \leq 6.8$. Following their methods and code\footnote{\url{https://bitbucket.org/ShenXuejian/quasarlf/src/master/}} we include QLFs at $z \sim 0.2, 1.2, 2.2, 3.2, 4.2, 5.2, \mathrm{and} \; 6.2$.

We consider two different models for the BHMF, the first is the expected BHMF resulting from a convolution between the observed galaxy stellar mass function \citep[GSMF, ][]{Leja_2020} and the local scaling relation between BH mass and galaxy bulge mass \citep[\mmb, ][]{Kormendy_Ho_2013}. We refer to this BHMF and its associated fits as \mone. In the other case, we use a posterior BHMF that was inferred from the GWB spectrum. Specifically we use the Le11ev model from \citet{Matt_2026a} which includes positive redshift evolution of the \mmb\ relation. We refer to this model as \mtwo. The relevant difference between these models is that \mtwo\ has a significantly increased number density of high-mass SMBHs relative to \mone\ (see Figure \ref{fig:bhmfs}). All posterior BHMFs that can reproduce the GWB spectrum require a significantly elevated number density of high-mass SMBHs relative to predictions from local galaxy and SMBH scaling relation predictions \citep[e.g.,][]{AgazieBHB_2023, Sato-Polito_2024, Liepold_2024, Matt_2026a}. The parameters for \mtwo\ come from fits to the best-fit 15 year Hellings and Downs (HD)-correlated free-spectrum representation of the GWB \citep{Hellings_1983, Lamb_2023}, also referred to as ``HD-w/MP+DP+CURN" in \citet{AgazieBHB_2023}. For this work we re-fit the Le11ev model to the most recent version of the GWB spectrum and used those posteriors for \mtwo\ \citep[see][for details on changes to the spectrum]{erratum}.

For all our fitting we used PyMC \citep{pymc_2023}. We sampled 100 chains each with 11,000 steps of which 1,000 were allocated for tuning. We used uniform priors on each parameter which are provided in Table \ref{tab:priors}. The low-\lamedd\ slope of the ERDF is highly uncertain. In our model we do not allow either of the ERDF slopes to be positive, but we do allow for flat slopes. The break point between slopes, \lbk, was allowed to vary across nearly the entire range of \lamedd\ values we considered and the upper bound was changed to 0.0 for the Eddington-limited case. We used these prior ranges across all redshifts and fit to each redshift bin independently; we discuss the behavior of each parameter with redshift in Section \ref{sec:results}.

\begin{deluxetable}{lll}
\tablewidth{0pt}
\tablehead{
\colhead{Parameter} & \colhead{Description}& \colhead{Prior Range}}
\startdata
\hline
\gone               & Lower slope of ERDF & $\mathcal{U}[-2.0, 0.0]$  \\
\gtwo               & Upper slope of ERDF & $\mathcal{U}[-5.0, 0.0]$ \\
$\log_{10}$ \lbk\   & Power law transition point & $\mathcal{U}[-3.0, 1.0]$ \\
\factive\           & Active fraction & $\mathcal{U}[0.0, 1.0]$ \\
\hline
\enddata
\caption{The prior ranges on the four free parameters in our model. All priors were uniform with wide ranges to allow for model flexibility. We use these same priors for the Eddington-limited case except for the upper bound of \lbk\ which we lower to 0.0 in that case.}
\label{tab:priors}
\end{deluxetable}

% with $\log M_\mathrm{Edd} = \log L - \log\lambda - \log C_\mathrm{Edd}$

% \begin{equation}
% \Phi_L(\log_{10} L_i) \approx \sum_j \tfrac12\big[g(\log_{10} L_i,\log_{10}\lambda_j) + g(\log_{10} L_i,\log_{10}\lambda_{j+1})\big]\,\Delta\log_{10}\lambda_j,
% \qquad g \equiv \Phi_\mathrm{BH},P,f_\mathrm{act}
% \end{equation}

\section{Results and Discussion}\label{sec:results}

\begin{figure*}[ht]\centering
    \includegraphics[width=\textwidth, keepaspectratio]{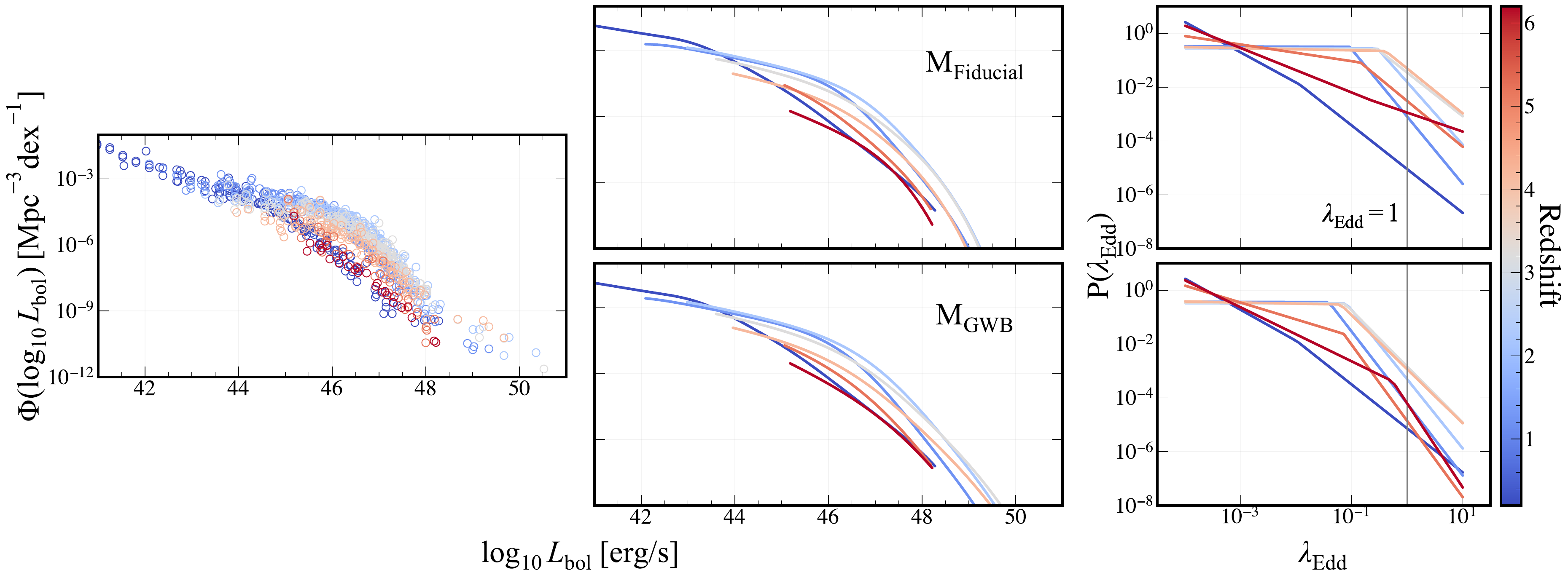}
    \caption{Here we show a summary of all of fits for \mone\ (top row,  middle panel) and \mtwo\ (bottom row, middle panel) and their corresponding ERDFs (right-most column). In the left-most panel we show the data from \citet{Shen_2020} that the models were fit to. The fits are only shown for the range of luminosities spanned by the data and everything is color-coded by redshift. The vertical gray line in the rightmost column represents the Eddington limit. Both models provide similarly good fits to the data, the only significant differences we see are for the highest luminosities $L_\mathrm{bol} \gtrsim 10^{48} \;\mathrm{erg} \; \mathrm{s}^{-1}$. The low-\lamedd\ portion of the ERDF is similar for both models, but the break between power laws, \lbk, is shifted towards lower values of \lamedd\ for \mtwo\ than \mone. The high-\lamedd\ slope is also steeper for \mtwo\ than \mone. Both models show similar redshift evolution of the ERDF, especially for \lbk.}
    \label{fig:fits_all_z_colorbar}
\end{figure*}

In this section we first describe the shape and quality of fit to the QLF data using our framework. We then describe the shape and behavior of the posterior ERDFs. Briefly, we find that we can produce realistic QLFs with our prescription for both \mone\ and \mtwo. We discuss individual redshifts in the following sections, while a summary of all our fits is presented in Figure \ref{fig:fits_all_z_colorbar}. Here we show all QLFs for one model in a single panel color-coded by redshift. The fits are only presented over the luminosity range of the data which are shown in the leftmost panel. For models that allow for super-Eddington accretion we see that both sets of fits have the same global shape and trends as the data. Except for $z = 1.2$, Bayes factors for these runs either show no preference between models or strongly favor \mtwo\ over \mone\ (see Table \ref{tab:posteriors}. With Eddington-limited runs, Bayes factors strongly favor \mtwo\ at all redshifts. We also find significant redshift evolution of the ERDF where the knee of the distribution increases with redshift until $z \sim 2$ after which it remains constant at $\log_{10} \lambda_\mathrm{break} = -0.55$ for \mone\ and -1.12 for \mtwo.

\subsection{Fits to the QLF}

\begin{figure*}[ht]\centering
    \includegraphics[width=\textwidth, keepaspectratio]{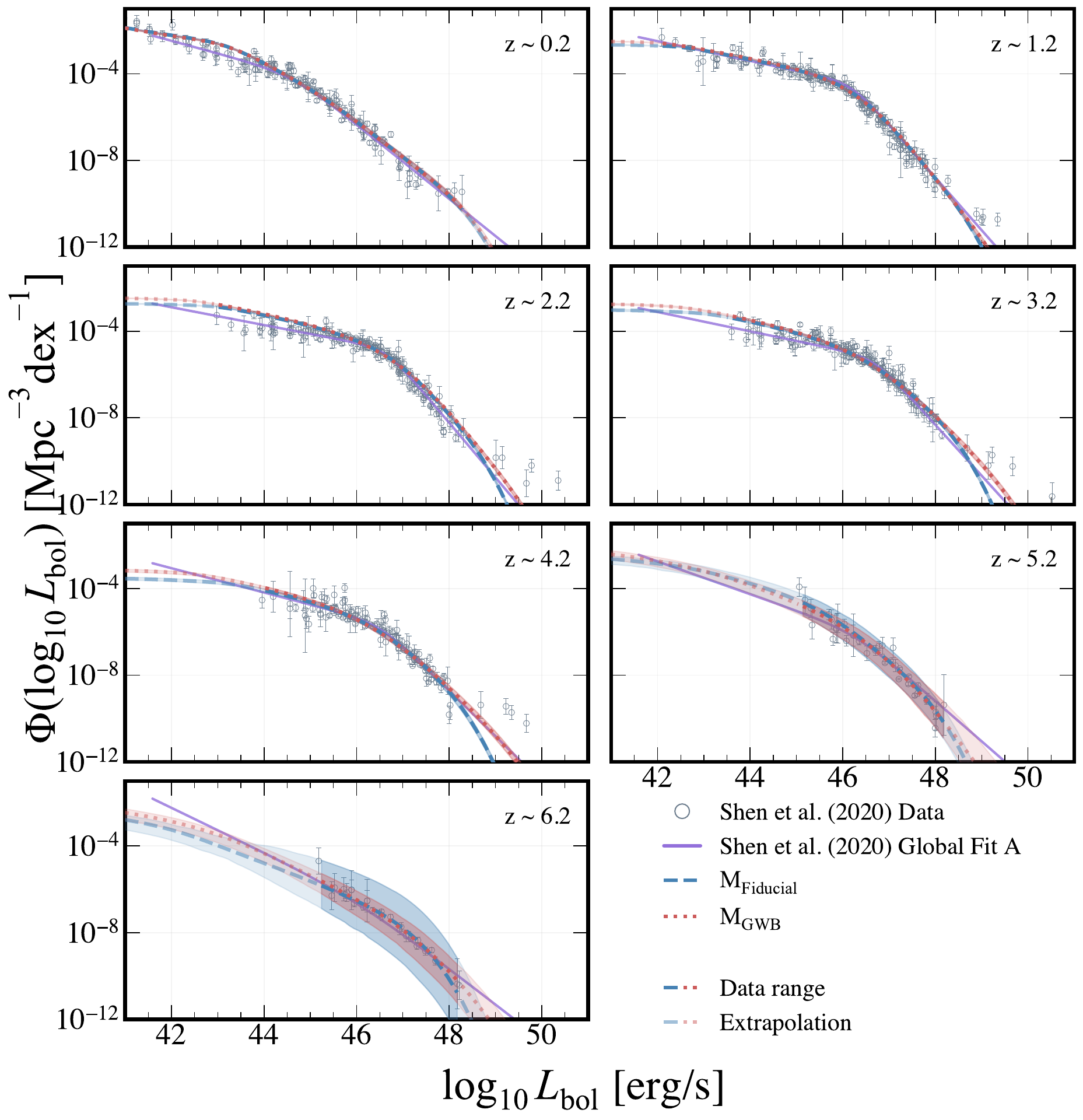}
    \caption{Here we show our fits from each model compared to the data and global fit A from \citet{Shen_2020}. The two models are generally equally good fits to the data, both models over-predict the number density for low luminosities for $z \sim 2.2  \;\mathrm{and}\; 3.2$. At the high-luminosity end, the \mtwo\ fits are able to match the data better than the \mone\ fits and produce similar number densities and shapes to to \citet{Shen_2020} fits.}
    \label{fig:qlf_fits_big_z_grid}
\end{figure*}

Our best-fit QLFs are compared to the data in Figure \ref{fig:qlf_fits_big_z_grid}. The posterior model for global fit A from \citet{Shen_2020} is also included for reference. The associated ERDFs and best-fit parameters are in Figures \ref{fig:erdf_fits_z_grid} and \ref{fig:params_vs_z_lbt} which we discuss in more depth in Section \ref{sec:erdf}. For all redshifts, both models provide good fits to the data and, in some cases, \mtwo\ is able to reach higher number densities at the high-luminosity end than with \mone. For $z \sim 0.2$, the best-fit ERDF and QLFs are only negligibly different between the two models, which is to be expected since their BHMFs are nearly identical at this point. At higher redshifts, the posterior ERDF for \mtwo\ returns a \lbk\ that is systematically lower than in \mone\ and the high-\lamedd\ slope, \gone\ is steeper. At low luminosities, both models show similar behavior to each other, at each redshift they either recover or over-predict the QLF number density below the characteristic luminosity. At high luminosities, especially for $L_\mathrm{AGN} > 10^{48}\; \mathrm{erg} \; \mathrm{s}^{-1}$, \mtwo\ consistently predicts more AGN than \mone; this difference is significant for $z \gtrsim 2.2$. This increased number density predicted from \mtwo\ is more in line with the data and is very similar to the fits from \citet{Shen_2020} in this luminosity range for $1.2 \leq z \leq  4.2$.

% \footnote{The data in the QLF from \citet{Shen_2020} have a maximum of $L_\mathrm{bol} = 10^{50.7}\; \mathrm{erg} \; \mathrm{s}^{-1}$. The bolometric corrections in \citet{Shen_2020} were applied by numerically inverting the relation in \citet{Hopkins_2007e} which can inflate the highest luminosities. The values above $L_\mathrm{bol} \sim 10^{48}\; \mathrm{erg} \; \mathrm{s}^{-1}$ should be treated with caution.}

Both of our model BHMFs start with a double Schechter shape which is inherited from the GSMF. The shape of the QLF is well described by a double power law. The power-law slope of the high-luminosity end of the QLF is shallow whereas Schechter functions drop off exponentially at high-masses/luminosities. The QLFs we obtain must match the shape of the observed QLF data only by changes to the ERDF. Even though our framework has few free parameters, the bulk of the data are captured by our model. The limitations of our methods are apparent in the extremes of the QLF fits. Flattening the slope of the high-luminosity end from the Schechter drop off can be done either by allowing for a significant fraction of the population to accrete above the Eddington limit \citep[\mone, also see][]{Aird_2013} or by having significantly more massive SMBHs (as with \mtwo). Between our two models, \mtwo\ does a better job recovering the flatter slope of the high-luminosity end of the QLF. For example in the $z \sim 3.2 \; \mathrm{and} \; 4.2$ panels of Figure \ref{fig:qlf_fits_big_z_grid} the fit for \mtwo\ (red dotted line) is higher and flatter than for \mone\ (blue dashed line) bringing it closer to the data (circles with error bars) in this area. \mone\ retains the sharp turn-down of the Schechter function at all redshifts. Schechter functions struggle to capture the sharp break that double-power laws have. Therefore reproducing the high-luminosity shape can come at the expense of fitting the low-luminosity end of the QLF. We see this effect in several redshift bins though the fits in this region are within the scatter of the data.

% Overall, our models are able to reliably fit the data though the extrapolated predictions significantly deviate from the double-power-law form assumed by \citet{Shen_2020}.

\subsection{Behavior of the ERDF}\label{sec:erdf}

\begin{figure*}[ht!]\centering
    \includegraphics[width=\textwidth, keepaspectratio]{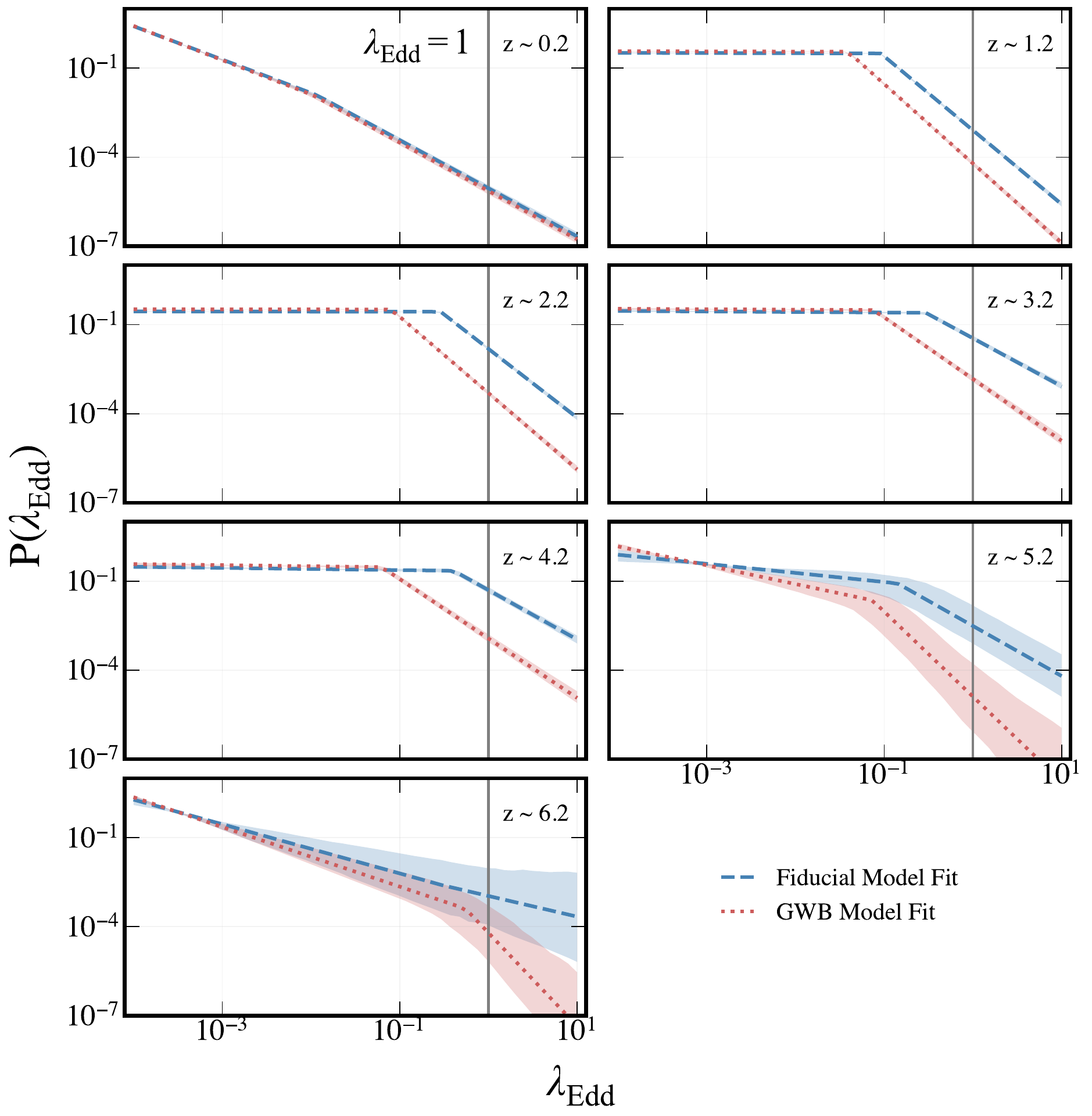}
    \caption{The ERDFs for \mone\ and \mtwo\ for the same redshifts as in Figure \ref{fig:qlf_fits_big_z_grid}. The vertical line marks the Eddington limit in all panels. The knee for \mtwo\ is shifted to lower values than in \mone\ for all $z > 0.2$. For the highest redshifts the high-\lamedd\ slope is steeper for \mtwo\ than \mone. The knee of the distribution increases with redshift for both models and the low-\lamedd\ slope flattens significantly for $1.2 \lesssim z \lesssim 4.2$.}
    \label{fig:erdf_fits_z_grid}
\end{figure*}

\begin{figure*}[ht]\centering
    \includegraphics[width=\textwidth, keepaspectratio]{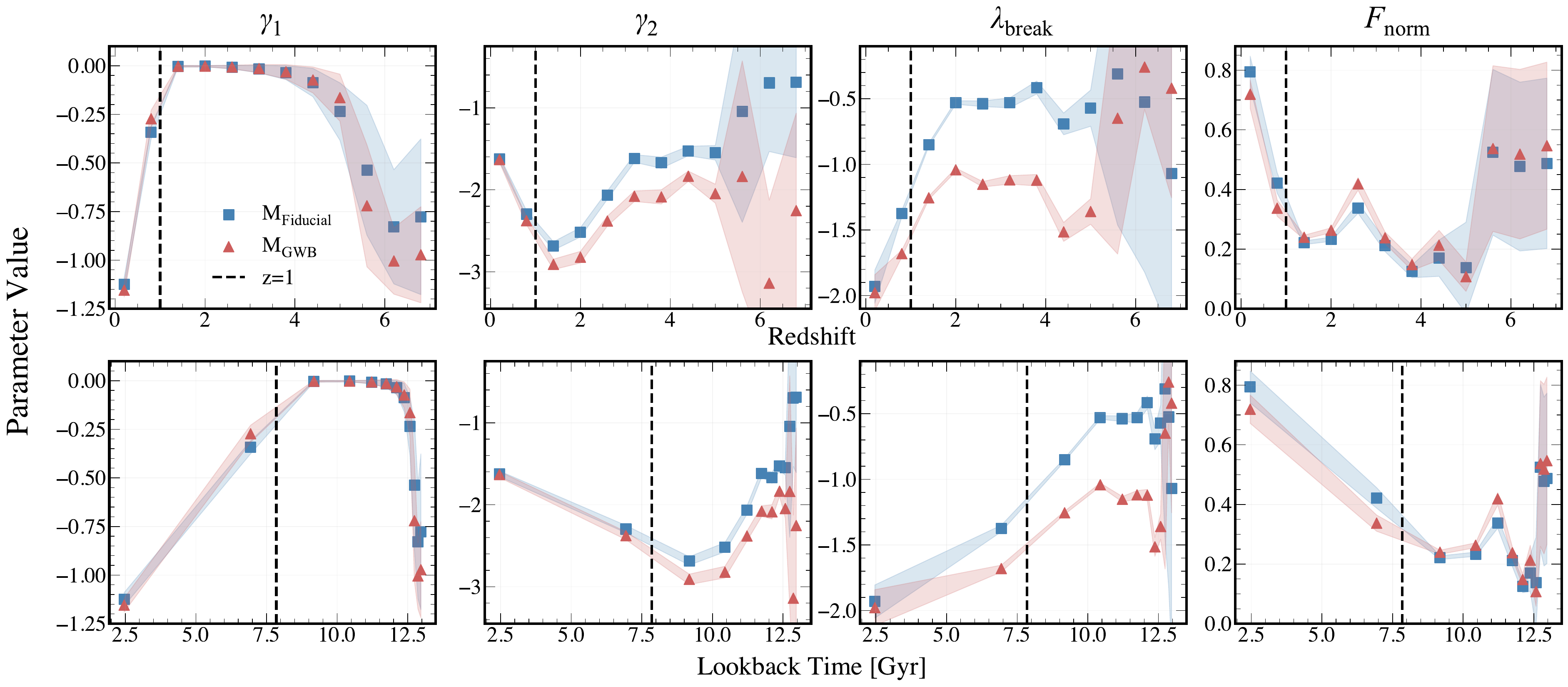}
    \caption{The best-fit parameters from our model versus redshift (top) and lookback time (bottom). The blue squares correspond to \mone\ and the red triangles are for \mtwo. The shaded regions represent the 1$\sigma$ confidence interval for each parameter at that redshift. The dashed black line indicates $z = 1$ in all panels. For \gone\ and \factive\ there are very few differences between the models. The best-fit values for \gone\ and \lbk\ show the same general trends with redshift for both models, but the values are are systematically lower for \mtwo\ than \mone\ at all redshifts for $0.2 < z < 5$. These two parameters control the steepness of the high-\lamedd\ slope and the break between power laws for the ERDF. Overall the fits for \mtwo\ require less extreme Eddington fractions to produce the observed QLF than \mone.}
    \label{fig:params_vs_z_lbt}
\end{figure*}

The posterior ERDFs are shown in Figure \ref{fig:erdf_fits_z_grid} and the best-fit parameters plotted against redshift and lookback time in Figure \ref{fig:params_vs_z_lbt}. Both posterior ERDFs show the same global behavior with redshift. All four parameters show rapid evolution for $ 0.2 \lesssim z \lesssim 2.0$. This evolution mimics the observed change in luminosity density across this redshift range, i.e.``cosmic downsizing''.

% Both models prefer rather flat values of \gone\ for $1.5 \lesssim z \lesssim 5$. We do not allow for \gone\ to be positive, the maximum of the prior was set to be zero. 

We see that, at intermediate redshifts, the models prefer relatively flat ($\gamma_1 \sim 0$) slopes. The flattening of the low-\lamedd\ potion of the ERDF has been seen by studies on X-ray data \citep[e.g., ][]{Schulze_2010, Schulze_2015, Bongiorno_2016, Li_2023, He_2024}. The flattening of the ERDF at low values of \lamedd\ may be a mass-dependent effect where the most massive SMBHs have flatter slopes \citep{Aird_2018, Zou_2024_bhar}. Alternatively, this flattening can be artificially induced by incomplete datasets for the dimmest AGN \citep{Kollmeier_2006, Shen_2012, Kelly_2013, Hickox_2014}. The QLF data and GSMF data that our BHMFs are based on are both highly complete so we do not believe this behavior is a completeness bias. Because of the way we normalize our ERDF, the \gone\ parameter is strongly negatively correlated with \factive\ (correlation coefficient between $-0.24$ and $-0.90$ depending on redshift) which is apparent by the reversed behavior with redshift for these two parameters. The shape and range of \lamedd\ we consider affect the total y-axis normalization of the resulting QLF. This effect does not impact our results for \gtwo\ or \lbk, but could induce non-physical slopes for the low-\lamedd\ portion of the ERDF.

There are notable differences between the best-fit parameters of our models for \gtwo\ and \lbk. Both of these parameters control the high-\lamedd\ end of the ERDF. For $z > 0.2$ the value of \gtwo\ is lower for \mtwo\ than for \mone. The slope of the high-\lamedd\ tail, \gtwo, describes the most extreme, but rare population of AGN that accrete above the break point of the distribution. Lower values of \gtwo\ disproportionately reduce the probability of an AGN to accrete at high- and super-Eddington rates. \gtwo\ decreases with redshift for both models until turning around at $z \sim 1.5$ after which it grows and levels off for $z \gtrsim 3$. With a systematically lower slope, the already rare population of the most extreme accretors becomes even rarer in \mtwo. It is interesting that both models minimize this parameter for $1 < z < 2$ since we would expect the fraction of super-Eddington accretion to correlate with luminosity density. This drop in \gtwo\ coincides with a rise in \lbk\ (which overall increases the probability of high- and super-Eddington accretion) and a rise in \gone, so the specific behavior of \gtwo\ parameter is largely due to its degeneracies with other parameters rather than a correlation with overall luminosity density (see also Figure \ref{fig:cornerplot}).

The break between slopes of the ERDF, \lbk, increases with redshift until $z \sim 2.0$ and then remains roughly constant for higher redshifts. The precise behavior for $z > 5$ is indiscernible from our analysis so it is unclear whether \lbk\ remains constant for the highest redshifts. After this point there are fewer than 50 data points in each redshift bin hence the fits are highly uncertain. For this parameter we see the most prominent differences between our two models. At all redshifts $0.2 < z < 5.0$, \lbk\ is lower for \mtwo\ than for \mone\ by factor of 0.58 on average. The value of \lbk\ shifts the entire ERDF along the x-axis with lower values of \lbk\ decreasing the probability for all $\lambda_\mathrm{Edd} > \lambda_\mathrm{break}$. Therefore the overall population of AGN implied by \mtwo\ is accreting at lower Eddington ratios than in \mone.

Previously, also creating QLFs from galaxy populations, \citet{Aird_2013} found that high values of \lbk\ are needed to reproduce the high-luminosity tail of the QLF. While neither model finds values of \lbk\ as high as \citet{Aird_2013}, the general trend is consistent. We also note that more recent studies have determined the value of \lbk\ to be securely below the Eddington limit and as low as $\lambda_\mathrm{break} = -1.5$ \citep{Zou_2024_bhar, Ananna_2022}. The knee of the ERDF has been observed to change with redshift and/or with mass \citep[e.g., ][]{Shen_2012, Bongiorno_2016, Aird_2018, Bernhard_2018, Ananna_2022, Zou_2024_bhar}. Given the quality of fits using our simple model, it appears that any true mass dependence is only second-order trend \citep[a result that is also recovered by some X-ray-based AGN studies][]{Aird_2018, Zou_2024_bhar, Igo_2026}.

The \factive\ parameter drops significantly until $z \sim 1$ after which it continues to decline, but less dramatically. This parameter tracks the fraction of all SMBHs that accrete above our lower limit of $\log_{10}(\lambda_\mathrm{Edd}) = -4$. We include SMBH masses as low as $M_\mathrm{BH} = 10^5 M_\odot$ which, with an Eddington fraction of $\log_{10}(\lambda_\mathrm{Edd}) \geq -4$ would have a luminosity of $L_\mathrm{AGN} \sim 10^{39} \; \mathrm{erg} \; \mathrm{s}^{-1}$ which is well below the lowest luminosity of the data, especially at higher redshifts. Therefore \factive\ tracks the theoretical fraction of all SMBHs that are active, but does not directly track the fraction of \textit{observable} AGN. The low-luminosity limit of the data rises roughly two orders of magnitude between $z = 0$ and $z = 1$ while we keep the same limits on the SMBH mass function. Therefore the total number of AGN in the QLF (as determined only over the range of the observed luminosity) is decreasing while the number of SMBHs remains constant. It is for this reason that the value of \factive\ drops with redshift.

Restricting our functions to the the observational limits (and known SMBH mass measurements) we can integrate the QLF over $43 < \log_{10} L_\mathrm{bol} < 51$  and the BHMF $6 < \log_{10} M_\mathrm{BH} < 10$. We find that, the fraction across redshifts ranges from 0.014 to 0.19 with an average of 0.057. This calculation assumes that the active fraction is not mass- or redshift-dependent, but our result is broadly consistent with literature reports of the observed active fraction \citep[e.g.,][]{Goulding_2010, Ananna_2022, Zou_2024_bhar, Pizzati_2024b}.

\begin{deluxetable*}{ccccccc}
\tablewidth{0pt}
\tablehead{\colhead{Redshift} & \colhead{Model} & \colhead{$\gamma_1$} & \colhead{$\gamma_2$} & \colhead{$\lambda_\mathrm{break}$} & \colhead{\factive} & \colhead{$\ln \mathcal{B}(\mathrm{M}_\mathrm{GWB} / \mathrm{M}_\mathrm{Fiducial})$}}
\startdata
\hline
\multicolumn{7}{c}{Runs allowing Super-Eddington Accretion} \\
\hline
\hline
\twoline{0.2} & \mone & $-1.12 \pm 0.03$ & $-1.62 \pm 0.03$ & $-1.94 \pm 0.13$ & $0.79 \pm 0.05$ & \twoline{$-2.8$} \\
              & \mtwo & $-1.15 \pm 0.04$ & $-1.63 \pm 0.04$ & $-1.98 \pm 0.14$ & $0.72 \pm 0.05$ & \\
\hline
\twoline{1.2} & \mone & $-0.01 \pm 0.01$ & $-2.50 \pm 0.04$ & $-1.04 \pm 0.01$ & $0.20 \pm 0.01$ & \twoline{$-9.4$} \\
              & \mtwo & $0.00 \pm 0.01$ & $-2.67 \pm 0.05$ & $-1.41 \pm 0.01$ & $0.22 \pm 0.01$ & \\
\hline
\twoline{2.2} & \mone & $0.00 \pm 0.00$ & $-2.32 \pm 0.05$ & $-0.54 \pm 0.02$ & $0.30 \pm 0.01$ & \twoline{$566.4$} \\
              & \mtwo & $0.00 \pm 0.00$ & $-2.57 \pm 0.06$ & $-1.10 \pm 0.02$ & $0.35 \pm 0.01$ & \\
\hline
\twoline{3.2} & \mone & $-0.02 \pm 0.02$ & $-1.62 \pm 0.05$ & $-0.53 \pm 0.04$ & $0.21 \pm 0.02$ & \twoline{$696.2$} \\
              & \mtwo & $-0.01 \pm 0.02$ & $-2.08 \pm 0.06$ & $-1.12 \pm 0.03$ & $0.24 \pm 0.02$ & \\
\hline
\twoline{4.2} & \mone & $-0.04 \pm 0.05$ & $-1.67 \pm 0.07$ & $-0.39 \pm 0.05$ & $0.08 \pm 0.02$ & \twoline{$192.1$} \\
              & \mtwo & $-0.04 \pm 0.05$ & $-2.00 \pm 0.08$ & $-1.20 \pm 0.05$ & $0.11 \pm 0.02$ & \\
\hline
\twoline{5.2} & \mone & $-0.31 \pm 0.18$ & $-1.70 \pm 0.16$ & $-0.84 \pm 0.36$ & $0.63 \pm 0.25$ & \twoline{$2.1$} \\
              & \mtwo & $-0.63 \pm 0.20$ & $-2.83 \pm 0.80$ & $-1.14 \pm 0.27$ & $0.61 \pm 0.26$ & \\
\hline
\twoline{6.2} & \mone & $-0.83 \pm 0.27$ & $-0.71 \pm 0.84$ & $-0.66 \pm 1.29$ & $0.48 \pm 0.28$ & \twoline{$-0.9$} \\
              & \mtwo & $-1.00 \pm 0.18$ & $-3.12 \pm 1.03$ & $-0.27 \pm 0.37$ & $0.53 \pm 0.28$ & \\
\hline
\hline
\multicolumn{7}{c}{Eddington-limited runs} \\
\hline
\hline
\twoline{0.2} & \mone & $-1.27 \pm 0.07$ & $-0.56 \pm 0.61$ & $-0.46 \pm 0.98$ & $0.93 \pm 0.08$ & \twoline{$26.1$} \\
              & \mtwo & $-1.24 \pm 0.08$ & $-1.36 \pm 0.60$ & $-1.98 \pm 0.98$ & $0.82 \pm 0.10$ & \\
\hline
\twoline{1.2} & \mone & $-0.01 \pm 0.01$ & $-1.98 \pm 0.05$ & $-1.12 \pm 0.02$ & $0.21 \pm 0.01$ & \twoline{$500.8$} \\
              & \mtwo & $0.00 \pm 0.01$ & $-2.57 \pm 0.06$ & $-1.42 \pm 0.01$ & $0.22 \pm 0.01$ & \\
\hline
\twoline{2.2} & \mone & $-0.01 \pm 0.01$ & $-0.85 \pm 0.08$ & $-0.64 \pm 0.04$ & $0.24 \pm 0.01$ & \twoline{$2711.1$} \\
              & \mtwo & $0.00 \pm 0.00$ & $-2.55 \pm 0.09$ & $-1.08 \pm 0.02$ & $0.33 \pm 0.01$ & \\
\hline
\twoline{3.2} & \mone & $-0.67 \pm 0.51$ & $0.00 \pm 0.16$ & $-2.63 \pm 0.51$ & $0.31 \pm 0.24$ & \twoline{$2966.6$} \\
              & \mtwo & $-0.02 \pm 0.02$ & $-1.77 \pm 0.09$ & $-1.15 \pm 0.05$ & $0.23 \pm 0.02$ & \\
\hline
\twoline{4.2} & \mone & $-0.96 \pm 0.52$ & $0.00 \pm 0.12$ & $-2.53 \pm 0.46$ & $0.37 \pm 0.27$ & \twoline{$1285.0$} \\
              & \mtwo & $-0.05 \pm 0.06$ & $-1.82 \pm 0.13$ & $-1.18 \pm 0.07$ & $0.10 \pm 0.02$ & \\
\hline
\twoline{5.2} & \mone & $-0.67 \pm 0.48$ & $-0.03 \pm 0.26$ & $-2.18 \pm 0.76$ & $0.43 \pm 0.27$ & \twoline{$124.3$} \\
              & \mtwo & $-0.59 \pm 0.21$ & $-2.44 \pm 0.87$ & $-1.26 \pm 0.35$ & $0.62 \pm 0.25$ & \\
\hline
\twoline{6.2} & \mone & $-0.67 \pm 0.46$ & $-0.02 \pm 0.24$ & $-2.19 \pm 0.72$ & $0.46 \pm 0.27$ & \twoline{$293.1$} \\
              & \mtwo & $-1.00 \pm 0.22$ & $-1.07 \pm 1.18$ & $-0.56 \pm 0.82$ & $0.48 \pm 0.28$ & \\
\hline
\enddata
\caption{The median posterior fit values and $1\sigma$ confidence intervals for the ERDF at seven redshifts for both of our models. The reported uncertainties are the output from the PyMC fitting code which incorporated the uncertainties on the QLF data, but not the input BHMFs. The right-most column shows the Bayes Factors in favor of \mtwo\ versus \mone\ for the QLF fits at each redshift. The top half of the table corresponds to the case where the ERDF is defined across $-4 < \log_{10} \lambda_\mathrm{Edd} < 1$ and the lower half is for the Eddington-limited case, $-4 < \log_{10} \lambda_\mathrm{Edd} < 0$. For the models that allowed for super-Eddington accretion the Bayes factors indicate either strong preference for \mtwo\ over \mone\ or no significant preference between the models, except for $z = 1.2$ where \mone\ is favored. In the case of the Eddington-limited runs, \mtwo\ is always decisively favored over \mone. It is worth noting that the data become sparse for $z \geq 5.2$ so the posterior distributions are poorly constrained, therefore the Bayes factors for these redshifts are more prior dependent and less reliable than the lower redshifts.}
\label{tab:posteriors}
\end{deluxetable*}

\subsection{Eddington-Limited Growth}\label{sec:eddlim}

\begin{figure*}[ht]\centering
    \includegraphics[width=\textwidth, keepaspectratio]{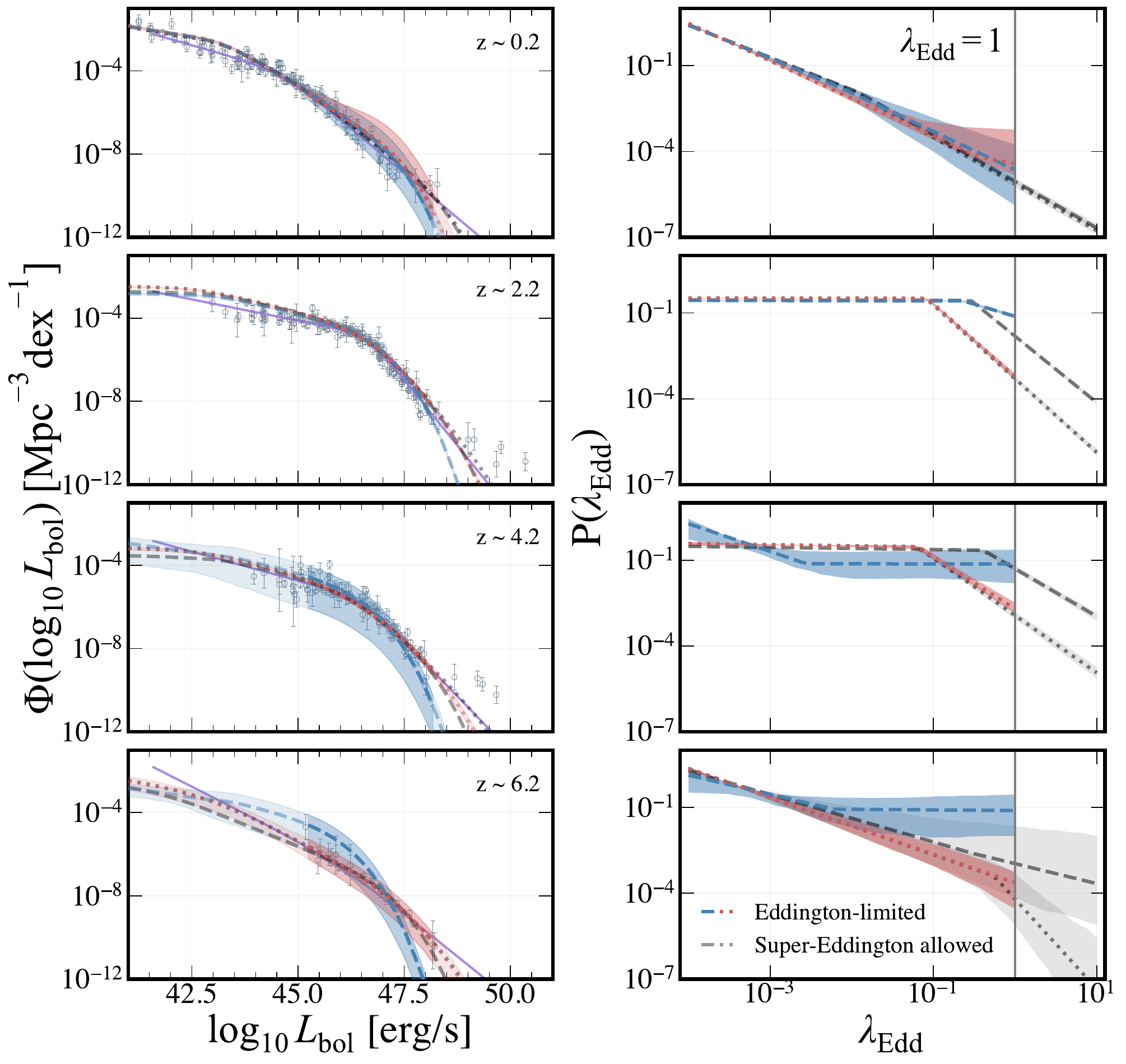}
    \caption{The results from our Eddington-limited runs. This figure uses the same style conventions as Figures \ref{fig:qlf_fits_big_z_grid} and \ref{fig:erdf_fits_z_grid}. Confidence intervals are not shown here for the QLF fits with allowed super-Eddington accretion for clarity. The results from Figures \ref{fig:qlf_fits_big_z_grid} and\ref{fig:erdf_fits_z_grid} are repeated as the low-opacity gray lines. Both best-fit QLFs drop off at the high-luminosity end compared to those with super-Eddington accretion allowed. The turn down is stronger for \mone\ while \mtwo\ maintains high number densities for the highest luminosities. We also see that the ERDF for \mone\ is significantly changed, especially at $z \gtrsim 4.2$ where \gone\ becomes flat apparently inverting the ERDF shape to maximize the relative probability of high-Eddington fractions. Conversely the ERDF for \mtwo\ is roughly identical to before, but now cut off at the Eddington limit. There is a minor turn up of \gone\ for both $z \sim 0.2$ and $z \sim 6.2$, but neither are significant. These results reinforce the general trend we see where \mone\ requires significantly higher probabilities of high- and super-Eddington fractions to match the QLF data than \mtwo.}
    \label{fig:el_fits}
\end{figure*}

Up until now we have defined the ERDF across $-4 < \log_{10}(\lambda) < 1$ meaning that we have allowed for accretion at up to 10 times the Eddington limit. We now consider the case of Eddington-limited growth and define the ERDF across $-4 < \log_{10}(\lambda) < 0$. These results are presented in Figure \ref{fig:el_fits} for four representative redshifts. The non-Eddington-limited fits for each model are repeated in gray for easy comparison.

Compared to the super-Eddington fits, both models have a notable drop in number density at the highest luminosities. This change is more extreme for \mone\ than \mtwo. For \mone\ the high-\lamedd\ tail of the ERDF begins to lift up for $z \gtrsim 2.2$ and \gtwo\ becomes flat for the highest redshifts. This flattening is a result of the models maximizing the relative probability of the highest Eddington fractions. We do not allow \gtwo\ to be positive so this value is at the maximum of the prior. Despite this drastic change to the ERDF the associated QLF for \mone\ falls short of the high-luminosity data even more so than before.

The best-fit QLF for \mtwo\ also lessens at the high-luminosity end compared to before, but to a lesser extent. The ERDF for \mtwo\ at each redshift appears nearly identical to the previous fit as though it were simply truncated at the Eddington limit. There are minor deviations for \gtwo\ at $z \sim 0.2\; \mathrm{and}\; 6.2$, but they are consistent everywhere to within one standard deviation.

These results reinforce our finding that, unlike with \mone, we can reproduce the QLF using \mtwo\ without requiring significant super-Eddington accretion.

\section{Physical Implications and Broader Discussion}\label{sec:discussion}

In this study we present a model that provides a good fit to both the NANOGrav GWB free spectrum and QLFs at $0.2 \leq z \leq 6.2$. This model also maintains consistency with the best constraints for the GSMF and local \mmb\ scaling relation. Our work is the first to use fits to both the GWB spectrum and QLF data and also the first to consider multiple frequency bins of the GWB spectrum in such a comparison. Our results imply the existence of a significant population of massive SMBHs that sit predominantly above the \mmb\ relation at high redshift and accrete relatively slowly throughout cosmic time.

Previous constraints on the BHMF vary significantly depending on the methods and assumptions of the studies. In particular, results from AGN studies typically report lower number densities of SMBHs than those inferred from the GSMF and local \mmb\ relation \citep{Shankar_2016, Liepold_2024}. Both of these approaches fail to produce the number densities of SMBHs necessary to explain the GWB \citep{AgazieBHB_2023, Matt_2026a}. We are able to resolve these tensions by altering two main assumptions. First, we use the posterior BHMFs from \citet{Matt_2026a} who allowed for the \mmb\ normalization to evolve positively with redshift. This evolution enables large GWB amplitudes without violating constraints for galaxy number density or local scaling relations. Second, as the focus of this paper, we fit for an ERDF that unifies the BHMF implied by the GWB with the observed QLF.
% Our posterior ERDF is shifted towards lower Eddington fractions than the fiducial model and shows significant redshift dependence that follows the same pattern reported by previous AGN studies \citep[e.g.,][]{Aird_2018, Zou_2024_bhar}.

To simultaneously model the SMBH population implied by the GWB (\mtwo) and the observed AGN populations, we find that the ERDF must be systematically shifted towards lower Eddington fractions than our fiducial model (\mone). Furthermore, the fits from \mone\ can only reproduce the high-luminosity end of the QLF if we allow for super-Eddington accretion. Even with allowing for \lamedd\ as high as 10 times Eddington, the QLF fits from \mone\ fall short of the data for $L_\mathrm{bol} \gtrsim 10^{47} \;\mathrm{erg} \; \mathrm{s}^{-1}$. Conversely, the QLF fits from \mtwo\ can reproduce the high-luminosity end of the QLF better than \mone\ and require only negligible super-Eddington accretion. The double power law model for the ERDF produces realistic QLFs and we find that the relative slopes and location of the break change significantly with cosmic time. The trends in each ERDF parameter are similar between models though the value of \gone\ and \lbk\ are systematically lower for \mtwo\ than \mone\ for most redshifts.

It is common practice to measure the specific ERDF which normalizes the Eddington fractions to galaxy masses \citep[e.g.][]{Aird_2013, Aird_2018, Zou_2024_bhar, Igo_2026}. This is informative and convenient because it eliminates uncertainty due to SMBH mass measurements and is not sensitive to the normalization of the assumed \mmb\ relation under the assumption that it is constant with redshift. If the \mmb\ relation is to be changing significantly with time, as may be implied by the GWB and high-redshift observations, then the specific ERDF would only be reliable for low redshifts. With a known \mmb\ normalization evolution, it would be possible to apply a correction factor to previous results, which would induce a redshift-dependent evolution of the ERDF towards lower \lamedd\ values.

Previous studies have concluded that the high-luminosity slope of the QLF require a BHMF that is shallower at high masses than implied by the GSMF \citep{Schulze_2010, Kelly_2013, Nobuta_2012}. The high-mass end of the GSMF drops dramatically because AGN activity impedes star formation and therefore inhibits galaxy stellar mass growth at the most extreme masses. This effect implies that the QLF must not be significantly impeded at the high-luminosity end assuming a broad general correlation between AGN luminosity and SMBH mass. The probabilistic nature of Eddington fractions across the AGN population partly solves this shape mismatch, but our results suggest that an increase to the overall number density of the most massive SMBHs (with or without a corresponding increase to the number of massive galaxies) can aid in this shape difference as much or more.

The results of \mtwo\ indicate that the high-luminosity end of the QLF is dominated by a population of extremely massive SMBHs accreting at sub-Eddington rates. This conclusion is not novel, for example \citet{Zhang_2024} determined that the highest luminosity quasars have masses higher than implied by the local \mmb\ relation and that low-redshift quasars primarily consist of the most massive, but sub-Eddington SMBHs. Furthermore, \citet{Caplar_2015, Caplar_2018} determined that the behavior of the characteristic luminosity of the QLF (the knee of the distribution) can be explained by assuming an $M_\mathrm{BH} / M_\star$ ratio that evolves positively with redshift. Additionally, \citet{Pizzati_2024a} investigated quasar halo clustering and found that, for $z \approx 4$, the AGN must be overmassive relative to the local \mmb\ relation and/or be in a state of high accretion.

The lower Eddington fractions in \mtwo\ imply that SMBHs are accreting at lower rates than \mone. This means that the total mass build up via accretion is reduced (for $0.2 \leq z \leq 6.2$). To grow these SMBHs within a Hubble time we would need need high seed masses, numerous mergers, and/or early accelerated growth. The merger rate of SMBHs is highly uncertain and binary hardening may occur on timescales of anywhere from several Myr to several Gyr, but recent EM-based and GWB-based advancements have suggested that the merger rates may be significant \citep{AgazieBHB_2023, Tahmasebzadeh_2026, Harris_2026, Dattathri_2025, Becsy_2026, McDonald_2026}. Furthermore, results from high-redshift EM observations and cosmological simulations support high seed masses and/or early accelerated growth establishing high SMBH masses early on, often followed by slower, self-regulating growth for $z < 5$ \citep{Kelly_2013, Alexander_2014, Lai_2024, Jeon_2025, Luberto_2025, Wu_2025, Hu_2025, Bhowmick_2026, Chen_2026, Naidu_2026}. Recently \citet{Pucha_2026} analyzed a sample of over 1.6 million line emitting galaxies at $z < 0.45$ from the first Dark Energy Spectroscopic Instrument (DESI) data release. They found two primary populations within their data; one group of overmassive SMBHs with low accretion rates and a separate group of normal-mass ratio SMBHs with higher accretion rates (relative to the other sample). They suggest that there are multiple coevolution pathways available to SMBH-galaxy pairs which produces these two populations and also inflates the scatter of the observed \mmb\ relation \citep[see also][]{Zhuang_2023, Cohn_2025, Matt_2026b}. Since the GWB is dominated by the most extreme SMBH masses \citep{AgazieBHB_2023}, it may be the case that the GWB constraints are primarily probing this extremely massive, but slowly accreting, population.

Our results are consistent with previous multimessenger studies of similar focus. By comparing the results from four different cosmological simulation suites \citet{Tillman_2026} concluded that the GWB amplitude was highly sensitive to accretion model choice and found a tension between feedback models that produced SMBH populations consistent with the GWB for $f_\mathrm{GW} = 1/1\mathrm{yr}$ versus those that could match galaxy observations. They concluded that the best model to unify all observations includes efficient SMBH growth early on. Using a semi-analytic model in combination with an N-body cosmological simulation, \citet{Bonoli_2025} determined that the GWB amplitude is best described by either heavy SMBH seeds or numerous light seeds accreting at super-Eddington rates. In this context, our sub-Eddington accretion rates for \mtwo\ would imply that these SMBH originate from heavy seeds though we cannot rule out models with super-Eddington accretion at the highest redshifts they consider ($z \sim 10$).

Future data releases from current PTA collaborations in addition to future GWB missions such as the \textit{Laser Interferometric Space Antenna} (LISA) will improve constraints on the BHMF. LISA will be particularly important for constraining moderate and intermediate mass SMBHs, which can be AGN, but are otherwise invisible to PTAs \citep[$10^5 < M_\mathrm{BH} < 10^7 M_\odot$][]{Sesana_2021, AgazieBHB_2023}.

\section{Summary}\label{sec:summary}

We present a simple model to convert between QLFs and BHMFs using a double-power law ERDF. We apply this model to a BHMF derived from the GSMF and local \mmb\ relation (\mone) as well as one inferred from the GWB (\mtwo). We are able to reproduce the observed bolometric QLF using both BHMFs with notable differences between the best-fit ERDFs. Compared to the results using \mone, the \mtwo\ model prefers lower Eddington fractions and produces better fits to the high-luminosity end of the QLF ($ L_\mathrm{bol} \gtrsim 10^{47} \;\mathrm{erg} \; \mathrm{s}^{-1}$). For both models, the break between power laws for the ERDF increases with redshift until $z \sim 2$ and remains roughly constant for higher redshifts. Our fits using \mtwo\ simultaneously match the GWB spectrum and QLFs for $0.2 \leq z \leq 6.2$ while maintaining consistency with galaxy and local SMBH observational constraints.

%% Please use the acknowledgment and contribution environments. This will 
%% be anonomyized when the "anonymous" style option is used.
\begin{acknowledgments}
C.M. is grateful for support from the University of Michigan Rackham Predoctoral Fellowship and computational resources from the Texas Advanced Computing Center at the University of Texas at Austin. The authors would like to thank Fan Zou for insightful discussions about this paper.

Anishinaabeg gaa bi dinokiiwaad temigad manda Michigan Kichi Kinoomaagegamig. Mdaaswi nshwaaswaak shi mdaaswi shi niizhawaaswi gii-sababoonagak, Ojibweg, Odawaag, minwaa Bodwe’aadamiig wiiba gii-miigwenaa’aa maamoonjiniibina Kichi Kinoomaagegamigoong wi pii-gaa aanjibiigaadeg Kichi-Naakonigewinning, debendang manda aki, mampii Niisaajiwan, gewiinwaa niijaansiwaan ji kinoomaagaazinid.  Daapanaming ninda kidwinan, megwaa minwaa gaa bi aankoosejig zhinda akiing minwaa gii-miigwewaad Kichi-Kinoomaagegamigoong aanji-daapinanigaade minwaa mshkowenjigaade.

The University of Michigan is located on the traditional territory of the Anishinaabe people. In 1817, the Ojibwe, Odawa, and Bodewadami Nations made the largest single land transfer to the University of Michigan.  This was offered ceremonially as a gift through the Treaty at the Foot of the Rapids so that their children could be educated. Through these words of acknowledgment, their contemporary and ancestral ties to the land and their contributions to the University are renewed and reaffirmed.
\end{acknowledgments}

% \begin{contribution}
% C.M. was responsible for the formal analysis, interpretation of results, and preparation of the manuscript. Both K.G. and R.H. served as mentors and aided in interpretation of the results and editing of the manuscript.
% \end{contribution}

% \facilities{HST(STIS), Swift(XRT and UVOT), AAVSO, CTIO:1.3m, CTIO:1.5m, CXO}


\software{Astropy \citep{astropy:2013, astropy:2018, astropy:2022} -- SciPy \citep{Scipy_2020} -- NumPy \citep{numpy} -- Matplotlib \citep{matplotlib} -- holodeck \citep{AgazieBHB_2023} -- PyMC \citep{pymc_2023}}


\appendix

\section{Appendix information}

\begin{figure*}[ht]\centering
    \includegraphics[width=\textwidth, keepaspectratio]{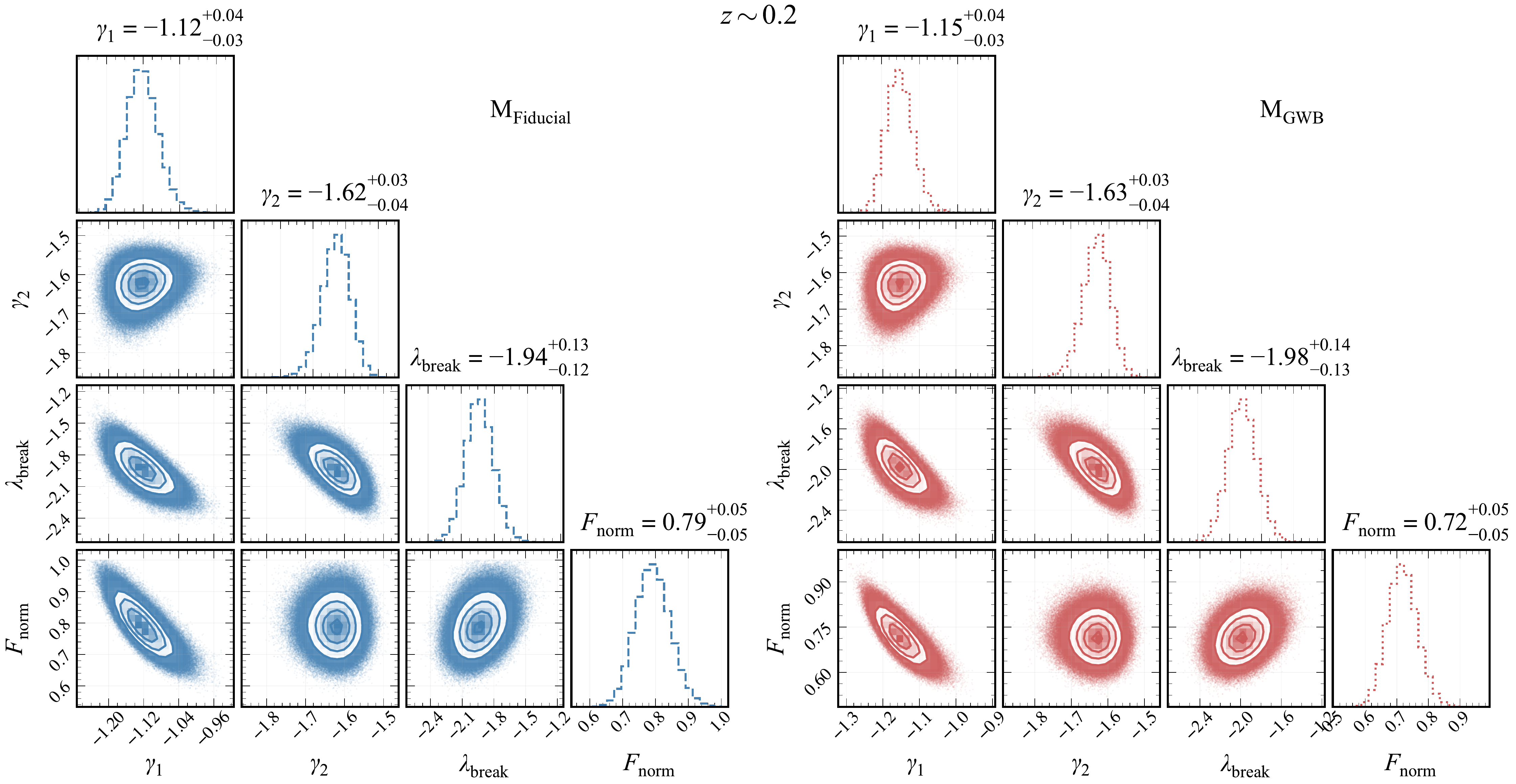}
    \caption{The corner plot for the lowest redshift fits for each of our two models. At this redshift the model posteriors are nearly identical. There are notable degeneracies between some parameters apparent by the non-circular shape to the contours. The strongest parameter degeneracy is between the \factive\ parameter and \gone. The complete figure set (14 images) is available in the online journal.}
    \label{fig:cornerplot}
\end{figure*}

\figsetstart
\figsetnum{1}
\figsettitle{Model Output Corner Plots}
\figsetgrpstart
\figsetgrpnum{1}
\figsetgrptitle{Corner plot for \mone\ and \mtwo\ at $z \sim 0.2$}
\figsetplot{cornerplot0.2.pdf}
\figsetgrpnote{All corner plots for our two models at each of the 7 redshifts between $0.2 \leq z \leq 6.2$. There are 14 total figures, 7 for the runs which allowed super-Eddington accretion and 7 for those that were Eddington-limited. The latter include "(Eddington-limited)" in the figure title.}
\figsetgrpend
\figsetgrpnum{2}
\figsetgrptitle{Corner plot for \mone\ and \mtwo\ at $z \sim 1.2$. Note that these two plots do not share axes between the same parameter panels.}
\figsetplot{cornerplot1.2.pdf}
\figsetgrpnote{All corner plots for our two models at each of the 7 redshifts between $0.2 \leq z \leq 6.2$. There are 14 total figures, 7 for the runs which allowed super-Eddington accretion and 7 for those that were Eddington-limited. The latter include "(Eddington-limited)" in the figure title.}
\figsetgrpend
\figsetgrpnum{3}
\figsetgrptitle{Corner plot for \mone\ and \mtwo\ at $z \sim 2.2$. Note that these two plots do not share axes between the same parameter panels.}
\figsetplot{cornerplot2.2.pdf}
\figsetgrpnote{All corner plots for our two models at each of the 7 redshifts between $0.2 \leq z \leq 6.2$. There are 14 total figures, 7 for the runs which allowed super-Eddington accretion and 7 for those that were Eddington-limited. The latter include "(Eddington-limited)" in the figure title.}
\figsetgrpend
\figsetgrpnum{4}
\figsetgrptitle{Corner plot for \mone\ and \mtwo\ at $z \sim 3.2$. Note that these two plots do not share axes between the same parameter panels.}
\figsetplot{cornerplot3.2.pdf}
\figsetgrpnote{All corner plots for our two models at each of the 7 redshifts between $0.2 \leq z \leq 6.2$. There are 14 total figures, 7 for the runs which allowed super-Eddington accretion and 7 for those that were Eddington-limited. The latter include "(Eddington-limited)" in the figure title.}
\figsetgrpend
\figsetgrpnum{5}
\figsetgrptitle{Corner plot for \mone\ and \mtwo\ at $z \sim 4.2$. Note that these two plots do not share axes between the same parameter panels.}
\figsetplot{cornerplot4.2.pdf}
\figsetgrpnote{All corner plots for our two models at each of the 7 redshifts between $0.2 \leq z \leq 6.2$. There are 14 total figures, 7 for the runs which allowed super-Eddington accretion and 7 for those that were Eddington-limited. The latter include "(Eddington-limited)" in the figure title.}
\figsetgrpend
\figsetgrpnum{6}
\figsetgrptitle{Corner plot for \mone\ and \mtwo\ at $z \sim 5.2$. Note that these two plots do not share axes between the same parameter panels.}
\figsetplot{cornerplot5.2.pdf}
\figsetgrpnote{All corner plots for our two models at each of the 7 redshifts between $0.2 \leq z \leq 6.2$. There are 14 total figures, 7 for the runs which allowed super-Eddington accretion and 7 for those that were Eddington-limited. The latter include "(Eddington-limited)" in the figure title.}
\figsetgrpend
\figsetgrpnum{7}
\figsetgrptitle{Corner plot for \mone\ and \mtwo\ at $z \sim 6.2$. Note that these two plots do not share axes between the same parameter panels.}
\figsetplot{cornerplot6.2.pdf}
\figsetgrpnote{All corner plots for our two models at each of the 7 redshifts between $0.2 \leq z \leq 6.2$. There are 14 total figures, 7 for the runs which allowed super-Eddington accretion and 7 for those that were Eddington-limited. The latter include "(Eddington-limited)" in the figure title.}
\figsetgrpend
\figsetgrpnum{8} %%%%% Begin Eddlim runs
\figsetgrptitle{Corner plot for the Eddington limited runs for \mone\ and \mtwo\ at $z \sim 0.2$. Note that these two plots do not share axes between the same parameter panels.}
\figsetplot{cornerplot0.2_edd_lim.pdf}
\figsetgrpnote{All corner plots for our two models at each of the 7 redshifts between $0.2 \leq z \leq 6.2$. There are 14 total figures, 7 for the runs which allowed super-Eddington accretion and 7 for those that were Eddington-limited. The latter include "(Eddington-limited)" in the figure title.}
\figsetgrpend
\figsetgrpnum{9}
\figsetgrptitle{Corner plot for the Eddington limited runs for \mone\ and \mtwo\ at $z \sim 1.2$. Note that these two plots do not share axes between the same parameter panels.}
\figsetplot{cornerplot1.2_edd_lim.pdf}
\figsetgrpnote{All corner plots for our two models at each of the 7 redshifts between $0.2 \leq z \leq 6.2$. There are 14 total figures, 7 for the runs which allowed super-Eddington accretion and 7 for those that were Eddington-limited. The latter include "(Eddington-limited)" in the figure title.}
\figsetgrpend
\figsetgrpnum{10}
\figsetgrptitle{Corner plot for the Eddington limited runs for \mone\ and \mtwo\ at $z \sim 2.2$. Note that these two plots do not share axes between the same parameter panels.}
\figsetplot{cornerplot2.2_edd_lim.pdf}
\figsetgrpnote{All corner plots for our two models at each of the 7 redshifts between $0.2 \leq z \leq 6.2$. There are 14 total figures, 7 for the runs which allowed super-Eddington accretion and 7 for those that were Eddington-limited. The latter include "(Eddington-limited)" in the figure title.}
\figsetgrpend
\figsetgrpnum{11}
\figsetgrptitle{Corner plot for the Eddington limited runs for \mone\ and \mtwo\ at $z \sim 3.2$. Note that these two plots do not share axes between the same parameter panels.}
\figsetplot{cornerplot3.2_edd_lim.pdf}
\figsetgrpnote{All corner plots for our two models at each of the 7 redshifts between $0.2 \leq z \leq 6.2$. There are 14 total figures, 7 for the runs which allowed super-Eddington accretion and 7 for those that were Eddington-limited. The latter include "(Eddington-limited)" in the figure title.}
\figsetgrpend
\figsetgrpnum{12}
\figsetgrptitle{Corner plot for the Eddington limited runs for \mone\ and \mtwo\ at $z \sim 4.2$. Note that these two plots do not share axes between the same parameter panels.}
\figsetplot{cornerplot4.2_edd_lim.pdf}
\figsetgrpnote{All corner plots for our two models at each of the 7 redshifts between $0.2 \leq z \leq 6.2$. There are 14 total figures, 7 for the runs which allowed super-Eddington accretion and 7 for those that were Eddington-limited. The latter include "(Eddington-limited)" in the figure title.}
\figsetgrpend
\figsetgrpnum{13}
\figsetgrptitle{Corner plot for the Eddington limited runs for \mone\ and \mtwo\ at $z \sim 5.2$. Note that these two plots do not share axes between the same parameter panels.}
\figsetplot{cornerplot5.2_edd_lim.pdf}
\figsetgrpnote{All corner plots for our two models at each of the 7 redshifts between $0.2 \leq z \leq 6.2$. There are 14 total figures, 7 for the runs which allowed super-Eddington accretion and 7 for those that were Eddington-limited. The latter include "(Eddington-limited)" in the figure title.}
\figsetgrpend
\figsetgrpnum{14}
\figsetgrptitle{Corner plot for the Eddington limited runs for \mone\ and \mtwo\ at $z \sim 6.2$. Note that these two plots do not share axes between the same parameter panels.}
\figsetplot{cornerplot6.2_edd_lim.pdf}
\figsetgrpnote{All corner plots for our two models at each of the 7 redshifts between $0.2 \leq z \leq 6.2$. There are 14 total figures, 7 for the runs which allowed super-Eddington accretion and 7 for those that were Eddington-limited. The latter include "(Eddington-limited)" in the figure title.}
\figsetgrpend
\figsetend

%% For this sample we use BibTeX plus aasjournalv7.bst to generate the

\bibliography{ref}{}
\bibliographystyle{aasjournalv7}



\end{document}